\documentclass[utf8%,draft 
]{ajaz} % ajazmp.cls

\usepackage{multirow}
\usepackage{array}
\usepackage{physics}
\usepackage{breqn}
\usepackage{caption}
\usepackage{graphicx}

\usepackage{natbib}
\usepackage{aas_macros}

\usepackage{newunicodechar}
\usepackage{multirow}

\makeatother

\newcommand{\Msun}{\ensuremath{\rm M_\odot}}
\newcommand{\Lsun}{\ensuremath{\rm L_\odot}}

\newcommand{\kms}{\,\mbox{km}\,\mbox{s}^{-1}}

\def\AJAzlogo{ }

\begin{document}

{\English
%\Russian
\setcounter{page}{1}

\title{Wolf-Rayet stars -- what we know and what we don't}
\setaffiliation1{Astronomical Institute of the Czech Academy of Sciences, Fri\v{c}ova 298, 25165 Ond\v{r}ejov, Czech Republic}

%\setaffiliation2{Institute, Region, Country}

\setauthor{O.~V.}{Maryeva}{1}
\email{olga.maryeva@asu.cas.cz}
%\setauthor{A.~A.}{Author2}{2}

\rtitle{Wolf-Rayet stars}
\rauthor{O.~V.~Maryeva}

\abstract{Today, we have a sufficiently complete picture of what the Wolf-Rayet (WR) stars are. Predictions of stellar evolution theory are in a good agreement with their parameters, estimated from observational data using stellar atmospheres codes; predictions of population synthesis also agree well with number of known WR stars. This article provides an overview of the main historical milestones in the studies of WR stars, showing how we came to this understanding, and what questions are still unanswered. \\

%This paper is overview of key milestones in the study of WR stars

%Nowadays we understand quite well what Wolf-Rayet stars are, and they are no longer a hot topic in astrophysics. However they still are keystones for many directions of research, from theory of stellar atmospheres up to stellar populations in other galaxies and gravitational waves. Through the existence of  the Wolf-Rayet phenomenon WR stars are linked, besides massive stars, with low mass evolved stars.

%In my review about current state of WR stars research I would like to touch such questions as:
%- evolutionary pathways of WR stars – stripped stars or final stage of single star evolution?
%- WR, luminous blue variables and red supergiants – is there an evolutionary connection?
%- WR in low metallicity – what facts do we have today?
%- Era JWST - new discoveries for thoughtful analysis. 

%исторический обзор -- основные этапы в исследовании WR звёзд 

	
}

\maketitle

%%%%%%%%%%%%%%%%%%%%%%%%%%%%%%%%%%%%%%%%%%%%%%%%%%%%%%%%%%%%%%%%%%%%%%%%

\section{Introduction}\label{intro}

Wolf-Rayet (WR) stars are a class of objects identified on the basis of their spectral features. WR spectra show strong emission lines of helium, nitrogen, carbon and oxygen in different stages of ionization. The width of these lines reaches tens of angstroms, the central intensities are sometimes 10-20 times the intensity of the continuum spectrum. A prominent feature of WRs are two bumps at 4650\footnote{Sometimes blue WR bump is also called Bowen blend.}  and 5808~\AA, which are groups of closely located lines of N\,{\footnotesize II}, N\,{\footnotesize III}, C\,{\footnotesize III} and C\,{\footnotesize IV}, which are visible even in low-resolution spectra. The lines are formed in an extended atmosphere -- in the stellar wind, expelling at velocities of $10^2-10^3~\kms$.  WR stars are characterized by a high mass-loss rate of several $10^{-5}~\Msun yr^{-1}$ and by high temperatures ($T_{\rm eff} \gtrsim 30 000$~K). WR are the final stage in the evolution of massive stars (stars with an initial mass $\gtrsim25~\Msun$) before the core-collapse supernova explosion. 

With accumulation of observational data it became clear that emission spectra and WR bumps are intrinsic to objects of different natures. Therefore nowadays there is a clear distinction between WR stars and \textit{WR phenomenon}, which occurs when fast-moving, hot plasma is expanding around a hot star \citep{Grafener2011,Vink2015}. WR phenomenon may occur in the low mass stars after ejecting their outer layers in the planetary nebula phase and exposing their hot cores prior to the white-dwarf phase \citep{vanderHucht1981, Marcolino2007, Todt2009}.  Central  stars of planetary nebulae showing WR phenomenon in their spectra are denoted as [WR] \citep{vanderHucht1981}. Rare and still unique case of low mass object with WR phenomenon and the absence of a clearly detected circumstellar nebula is LAMOST J040901.83+323955.6 \citep{Maryeva2024}. It was selected as a WR star by \citet{Skoda2020}, but \citet{Maryeva2024} showed that  this object is $0.9~\Msun$ star caught in a rare transitional phase from post-AGB to CSPN. Another unique object with  WR phenomenon is IRAS 00500+6713, formed as a result of a merger of two white dwarfs \citep{Gvaramadze2019}. The WR phenomenon is also observed in young supernovae \citep{Gal-Yam2014}. 

As we know now, massive stars showing WR spectra  are also not a homogeneous class. They are split into two groups:  {\it very massive stars} and {\it classical WR stars}. 
{\it Very massive stars} ($M_{\rm ini}\ge100 \Msun$) experience strong stellar winds and show WR spectra already during core H-burning phase, when they are located on main sequence. All of them show signs of hydrogen on their surface  and are classified as WNh  objects \citep{Smith1996, Crowther2010, Martins2023}. 
{\it Classical WR stars} (usually just named  {\it WR stars})  stars are hydrogen-depleted objects that have evolved off the main sequence and suffered intense mass loss \citep{Crowther2007, Shenar2024}. 
{\it Classical WR stars} will be the subject of this review article.

The field of contemporary studies of WR stars is enormous and cannot be fully covered in a single small review like this one, so I'd like to recommend two other detailed reviews --  \citet{Crowther2007} and  \citet{Shenar2024} -- for better understanding what WR stars are. On the other hand, here I will concentrate more on the historical aspects of WR studies, on how we have arrived to our present day understanding of them. Therefore Section~\ref{sec:history} will cover the history of studies of WR stars, while Section~\ref{sec:problems} will highlight several topical problems and open questions about them. 

%%%%%%%%%%%%%%%%%%%%%%%%%%%%%%%%%%
\section{From discovery to understanding}\label{sec:history}

In 1867 year astronomers from Paris observatory Charles Wolf and Georges Rayet discovered three stars in Cygnus constellation those spectra significantly differ from other stars. In contrast with  other stars, where usually absorption lines are dominant in spectra, discovered objects showed strong emission lines \citep{WolfRayet1867}. The class of objects received its name in honor of the discoverers and gradually began to be replenished with new members. 
One hundred years after the discovery of WRs, by the end of 1960-s the number of known WRs in the Galaxy had reached over a hundred, totalling 127 \citep{Roberts1962, Smith1968}, while now is $\sim 700$ \citep{Rosslowe2015}. At the same time, in the late 1950-s middle 1960-s, work began to search for extragalactic WRs -- WRs in the Large Magellanic Cloud \citep{WesterlundRodgers1959, WesterlundSmith1964}. 
% !!! начались исследования по поиску экстрагалактических WR в БМО

Significant contribution to the understanding of the physics of WR stars was given by Carlyle~S. \citet{Beals1929}. Beals found that some lines in the spectra of WR stars have a P\,Cygni profile and  based on comparison with P\,Cygni profile lines in spectra of novae he suggested that WRs are surrounded by expanding envelopes \citep{Beals1929}.  Beals also found a difference between WR stars and novae: the P\,Cygni profiles in the WR stars do not change over time. This allowed to propose that an outflow from WR stars occurs continuously~\citep{Beals1929}. This was confirmed by  \citet{Chandrasekhar1934}, who developed a solid footing for interpreting P\,Cygni profiles as arising in expanding atmospheres.   \citet{Kosirev1934} used the diagnostics developed by Chandrasekhar to estimate the mass loss and maximum outflow velocity of a WR star and found, respectively, $\sim10^{-5}~\Msun yr^{-1}$ and $\sim1000~\kms$.

At the beginning of the 20th century, stellar spectroscopy developed in close cooperation with atomic physics. Investigating spectra of the star $\zeta$\,Puppis %Edward Charles 
Edward Pickering paid his attention to previously unknown lines 5411, 4541, 4200, 4100, 4026, 3924, 3858, 3813, 3782~\AA\ \citep{Pickering1897}. He interpreted them as another series  of hydrogen (besides Balmer) \citep{Pickering1901}. Line 4686~\AA, first discovered during a solar eclipse and often found in WR spectra, was also considered to be a hydrogen line \citep{Fowler1912}. Although \citet{Fowler1912} was able to obtain these spectral lines in the laboratory in 1912, during an experiment with a helium-filled tube, he considered their appearance to be a contribution from hydrogen impurity. It was not until 1913 that Niels Bohr's work \citep{Bohr1913} explained the nature of these lines as ionized helium He\,{\footnotesize II}, confirming the atomic model (see  \citet{Robotti} for historical review). 

In 1890 Pickering drew attention to the similarity of the spectra of WR stars and planetary nebula \citep{Pickering1891}. Due to this, in papers devoted to emission spectra of nebulae it is possible to find descriptions of spectra of WR stars. 
The lines at 4647, 4650, 5696, 5801 and 5812~\AA\ were identified with the carbon by  \citet{Wright1918}. Ira~S. Bowen, who found an interpretation of the ``nebulium'' lines as [O\,{\footnotesize III}], gave a detailed list of emission lines observed in nebulae, which includes an identification of almost all lines visible in the WRs \citep{Bowen1928}. In a 1934 paper  \citet{Bowen1934} suggests a significant role of the fluorescence for formation of the N\,{\footnotesize III}~4634,~46340 lines belonging to the WR bump.

Already the spectra of the first discovered WR stars showed a difference: the broad emission bands are located in different places \citep{WolfRayet1867, Huggins1890}. After the identification of all the main spectral lines, \citet{Beals1933} proposed the splitting of WR into two groups. This classification was approved in 1938, the International Astronomical Union (IAU) divided the spectra of WR stars into types WN and WC, depending on whether the spectrum was dominated by lines of nitrogen or carbon-oxygen respectively \citep{Beals1933, Swings1942}. WN stars are believed to show the hydrogen burning products via the CNO cycle, while WC stars reveal the helium burning products via the triple-$\alpha$ cycle. Based on the strength of the emission lines and line ratios, WN stars can be further classified into the spectral subtypes WN2 to WN11, and WC stars into the spectral subtypes WC4 to WC9 \citep{Smith1968,Smith1990,Smith1994,Smith1996, Crowther1998}. 
%(Smith 1968; Smith et al. 1990, 1994, 1996; Crowther et al. 1998). 
Also, there are transition types from WN to WC, which are called WN/C, whose spectra show strong emission lines of carbon and nitrogen simultaneously \citep{ContiMassey1989}. 

In 1933 Victor~A.  \citet{Ambartsumian1933} estimated He/H ratio for WR stars using intensities of H$\beta$ and He\,{\footnotesize II}~4686 lines and found  He/H$\gtrsim1.8$. Subsequent studies (\citet{Rublev1972He} and references therein) confirmed excess of helium.  \citet{Gamow1943}  suggested that the anomalous composition of WR stars was the result of nuclear processed material being visible on their surfaces. Despite these arguments in favor of far evolved status of WR stars, debates about it continued until the mid-1970s. There was alternative hypothesis, that WR are young and more massive than T\,Tau objects, before main sequence \citep{Sahade1958, Underhill1968}. It is important to mention the studies of Sergej~V. Rublev, in which he developed methods for determining the temperatures and luminosities of WR stars and estimated the hydrogen abundances \citep{Rublev1965,Rublev1970,  Rublev1972, Rublev1975}. These works played a significant role in the formation of the modern understanding of WR stars as far evolved massive objects. 

%--------------------------------------------------
\subsection{Development of numerical methods}

Contemporary progress in our understanding of hot stars and especially WR stars has been significantly defined by the progress of numerical modelling of stellar atmospheres, in turn enabled by the development of both physical approximations and numerical methods. 

Victor~V. Sobolev developed the theory of radiative transfer in moving media and applied it to the determination of physical conditions in the envelopes of WR, Be, novae, etc. type stars \citep{Sobolev1947, Sobolev1960}. He showed that due to the presence of a velocity gradient, photons of spectral lines are able to leave the deep layers of the atmosphere directly, avoiding the usual diffusion process; this significantly simplifies the study of the radiative equilibrium of moving atmospheres. This theory became the basis for the transition to modern quantitative calculations of the stellar wind.

WR stars are essentially hot stellar cores surrounded by extended, low-density atmospheres. Due to their high luminosity the radiation in the stellar wind dominates over collisional processes, and the models of WR's atmospheres should be calculated taking into account the deviations from local thermodynamic equilibrium (non-LTE). 
Straightforward iterative procedure for the solution of radiative transfer equations -- $\Lambda$ iteration -- fails to converge in typical non-LTE conditions, as in scattering dominated media of large optical
thickness, its convergence is extremely slow (see overview by  \citet{Atanackovic-Vukmanovic2004} and references therein). 

%The well-known most straightforward iterative procedure, so-called $\Lambda$ iteration, solves the problem equations in turn. However, this simple procedure fails to converge in typical non-LTE conditions. $\Lambda$  iteration follows the process of a particular scattering event and at each iteration step it corrects the solutions only within a unit optical path. Therefore, in scattering dominated media of large optical thickness, the convergence of $\Lambda$  iteration is extremely slow. 

% statistical equilibrium (SE)
% Radiative transfer (RT) 
%Rybicki (1972) has introduced the so-called ''core saturation method'' -> Scharmer (1981, 1984) New approach to non-LTE problems -> Hamann (1985, 1986) ''Diagonal operator'' 
%''approximate lambda operator'' 

%Auer \& Mihalas (1969) Complete linearization scheme -> Cannon (1973) Perturbations in Radiative-Transfer Theory -> Hillier (1990, 1991) Tridiagonal Newton-Raphson operator

% Several approaches have been proposed in early 70's as a way to numerically solve radiative transfer (RT) problems in non-LTE case:
%\begin{itemize}
%    \item ``complete linearization  method'', Auer \& Mihalas (1969) \citep{AuerMihalas1969};
%    \item ``core saturation method'',  Rybicki (1972) \citep{Rybicki1972};
%    \item ``perturbation technique'', Cannon (1973a,b) \citep{Cannon1, Cannon2}.
%\end{itemize}
Several approaches have been proposed in early 1970-s as a way to numerically solve radiative transfer (RT) problems in non-LTE case. The first fully self-consistent solution of RT problems for non-LTE  case  was achieved by \citet{AuerMihalas1969} who developed the complete linearization  method  based on the Newton--Raphson scheme to solve the set of discretized structural equations in a robust manner that opened a way for significant progress in the modelling of stellar atmospheres. 

A different approach is the ``core saturation method'', introduced by \citet{Rybicki1972} who implemented  a modification of the $\Lambda$ iteration method that made it practically usable for solving the RT and statistical equilibrium equations. In order to eliminate the cause of the slow convergence of simple $\Lambda$ iteration (poor conditioning of the equations arising from a large number
of scatterings), Rybicki proposed the elimination of scattering events in the line core as they do not contribute much to the transfer process. 

%''Perturbations in Radiative-Transfer Theory''
%Another %breakthrough in computational RT was made with 
Another approach is ``perturbation technique'' proposed by  \citet{Cannon1, Cannon2} who was the first to use the idea of 'operator splitting'  in radiative transfer computations. He replaced the full (exact) $\Lambda$  operator by a simplified one (so called approximate Lambda operator, or ALO) and computed a small error term made by this approximation by using a perturbation technique. Cannon's operator perturbation technique was an efficient way to combine computational advantages of approximate solutions with the accuracy of exact methods.

In 1980-s computational approximations to accelerate the $\Lambda$ iteration started to be used together with the operator perturbation technique introduced by  \citet{Cannon1, Cannon2} to simplify the direct solution, making the basis for the class of methods known as Accelerated Lambda Iteration. 
Wolf-Rainer \citet{Hamann1985, Hamann1986} adopted perturbation technique to approximate lambda operator, based on idea of ``core saturation'' and presented test calculations for a typical WR star atmosphere. This effort became the foundation for the PoWR code \citep{Hamann2003} that is still being actively developed and successfully applied for numerical modelling of various types of astrophysical objects.
Independently, D.~John \citet{Hillier1990} presented a method based on tridiagonal Newton-Raphson operator, complete linearization scheme  and perturbation technique for calculations of WN and WC atmosphere models. Over time, this method has grown into CMFGEN code \citep{Hillier1998}.

%Since the mid-1990s, radiative transfer codes designed for modeling extended expanding atmospheres have become a reliable instrument for studying hot stars with high mass-loss rates. %These codes made it possible to calculate the atmospheres of outstanding objects, such as O-supergiants, WR stars and even luminous blue variables (LBVs).
Since the mid-1990s, CMFGEN and PoWR have become a reliable instruments for studying hot stars with high mass-loss rates. Systematic study of Galactic WR stars  started with a series of works by Paul Crowther  titled ``Fundamental parameters of WR stars'' and devoted to the analysis of properties of nitrogen-rich WR stars (WN) based on optical and ultraviolet (UV) data and numerical modelling using CMFGEN code \citep{Crowther1995, Crowther1995a, Crowther1995b, Crowther1995c, Crowther1995d}. Advances in infrared (IR) spectroscopy also enabled studies of WR stars near the Galactic center \citep{Martins2007, Liermann2010}. 
Modeling of large sample of Galctic WN \citep{Hamann2006} and WC \citep{Sander2012} stars ($\approx126$ objects in total) allowed to obtain a homogeneous set of stellar and atmospheric parameters for them, to determine the correct location of WR stars in the  Hertzsprung--Russell (HR) diagram.
%Analysis of large sample of Galctic WN \citep{Hamann2006} and WC \citep{Sander2012} stars ($\approx126$ objects in total) allowed to determine the correct location of WR stars in the HR diagram. 

%The most common codes –- CMFGEN \citep{Hillier1998} and PoWR \citep{Hamann2003} -– have been used to determine the properties of the majority of the Galactic nitrogen-rich Wolf–Rayet (WN) stars \citep{Crowther1995,Hamann2006}, %(Crowther, Hillier \& Smith 1995; Hamann, Grafener \& Liermann 2006), 
%including WN stars near the Galactic Centre \citep{Martins2007, Liermann2010}. 
% (Martins et al. 2007; Liermann et al. 2010) 

PoWR code was widely used for studying extragalactic objects.
Among WR stars in nearby galaxies, the ones in Magellanic Clouds are most studied \citep{Bestenlehner2014, Hainich2014, Hainich2015}. % (Bestenlehner et al. 2014; Hainich et al. 2014, 2015). 
Sample of 74 WR stars belonged to M33 galaxy was analysed by Pritzkuleit (2020) \citep{Pritzkuleit2020}.  Sander et al. (2014) investigated late-type WN stars in M31 galaxy \citep{Sander2014}.  These studies extended our understanding of properties of WR stars in different galactic environments and metallicity-dependence of massive star winds.

\subsection{Evolution}

Modern view on WRs as products of evolution of massive stars was finally formed only in 1980s. Although by the 1980s WR stars were already recognised as descendants of massive OB stars, their exact evolutionary status with respect to the main sequence and other evolved massive stars was still an open question. WR stars were being proposed as possible progenitors of supernovae, and particularly the newly-discovered type Ib supernovae that lack hydrogen but are apparently associated with young massive stars. 

The start of systematic X-ray observations with space-borne X-ray telescopes in 1960s 
significantly advanced the theory of close binary stars and their interactions, and it led  Bohdan \citet{Paczynski1967} to the suggestion that WR stars are the products of binary evolution -- namely, mass exchange in a binary system. In his scenario,  a more massive star in binary system evolves faster and, upon exhaustion of hydrogen in the core, fills the Roche lobe and a rapidly loses its hydrogen envelope that outflows towards the companion. As a result, when the envelope is fully transferred, its hot helium core remains as a WR star with little hydrogen in the outermost layers \citep{Paczynski1973}. 

The hypothesis of Paczy\'{n}ski was well accepted and it naturally explained the anomalous chemical composition of WR stars as naked stellar cores enriched with  products of nuclear reactions. On the other hand, based on the similarity of Of and WN stars %in the Carina Nebula 
(Figure~\ref{spectra}) and on the first estimates of the mass loss rate for hot massive stars based on UV observations Peter \citet{Conti1975} suggested that the stellar wind of O stars is strong enough to remove the outer layers and uncover the stellar core, thus proposing a single star mass loss channel to form WR stars. In addition, brightest WR star, $\gamma$\,Vel, was inconsistent with Paczy\'{n}ski's hypothesis as, while being a binary, its orbit was clearly elliptical and its second component lacks any signs of previous mass exchange, both suggesting that there was no interaction between the components \citep{Conti2015}.

%%%%%%%%%%%%%%%%%%%%%%%%%%%%
\begin{figure}
	\centering
	\includegraphics[width=0.7\textwidth]{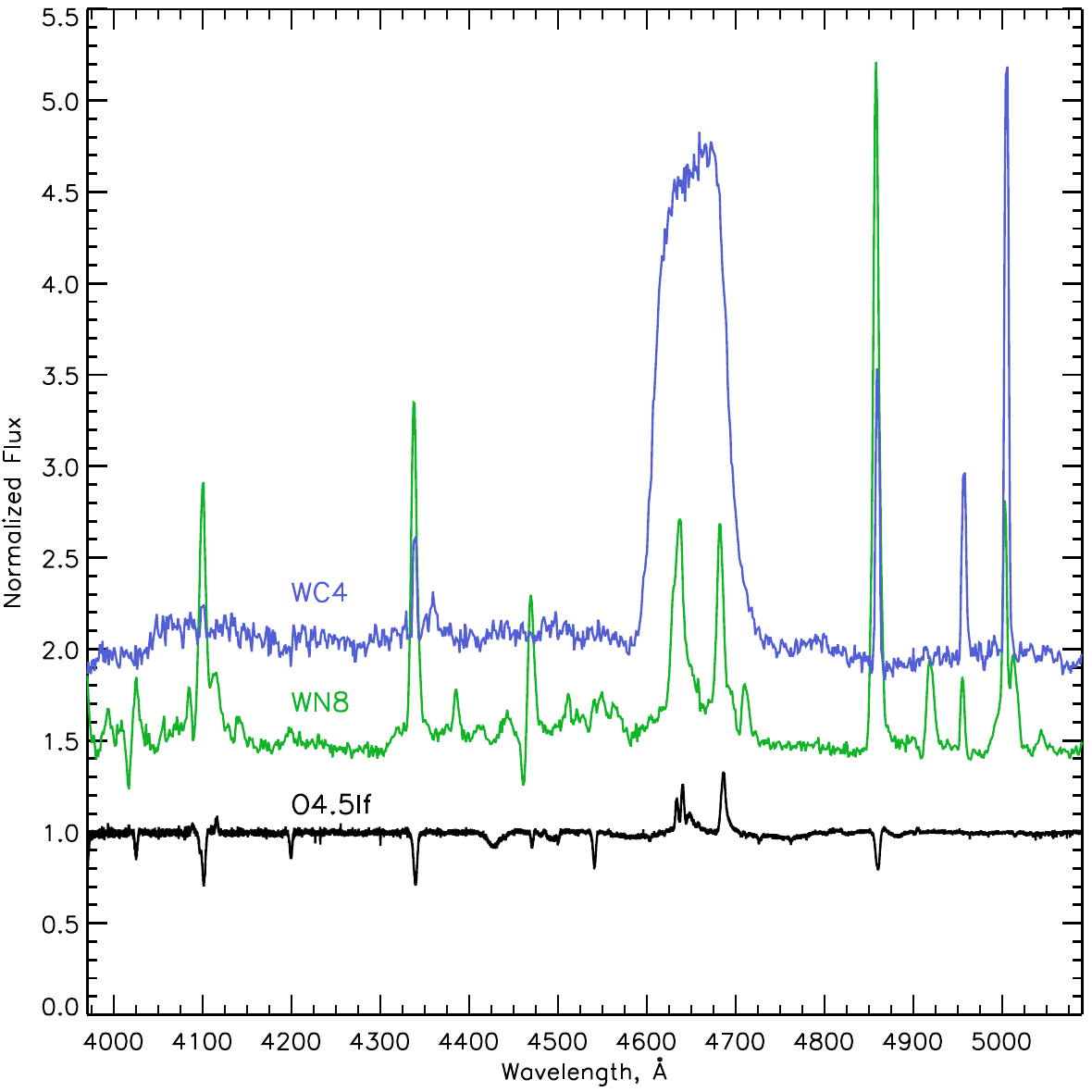}
	\caption{Comparison of spectra of J013340.19+303134.5 (WC4), J013352.43+304351.7 (WN8) and HD\,14947 (O4.5If) as illustration of significant differences in intensities of spectral lines for O stars and different sub-types of WR stars.}
	\label{spectra}
\end{figure}
%%%%%%%%%%%%%%%%%%%%%%%%%%%%%%%%%%

In 1984 Peter \citet{Conti1984} introduced a new class of objects. In his talk at the IAU symposium, Peter Conti grouped  S\,Doradus variables, Hubble–Sandage variables, $\eta$\,Carinae, P\,Cygni, and other similar stars together under the term ``luminous blue variables'' (LBV). 
LBV are characterized by strong mass loss (around $10^{-5}~\Msun yr^{-1}$) and occasional giant eruptive events \citep{Humphreys1994}.  Conti suggested that massive O stars transit to WR stars through this short-duration LBV phase. This idea is consolidated as the ``Conti scenario'' and becomes the main channel for WR star formation. Modern stellar evolution calculations are consistent with the ``Conti scenario'' \citep{Groh2014,Sander2019}. According to numerical calculations the stars with initial mass of $60~\Msun$ spend $2\times10^5$~yr in the LBV phase –- only 5\% of their lifetime \citep{Groh2014}. 
Recently observed transition from LBV to WN8 spectral type in Romano' star is also a direct confirmation of ``Conti scenario'' \citep{Polcaro2016, Maryeva2019}.

Similar scenario for WR formation was also proposed by  \citet{Bisnovatyi-Kogan1972}, with significant mass loss on the red supergiant phase. They show that large inverse density gradient appears in the outer layers of a $30~\Msun$ star in this phase due to supercritical luminosity, resulting in powerful and rapid mass loss and formation of a stripped helium core.

In the last decade, Paczy\'{n}ski's hypothesis has again been discussed as a mechanism for the formation of WR stars. Analysing the spatial positions of LBV and WR stars, Nathan Smith \citep{SmithTombleson2015, Smith2016,Smith2019}  
%(Smith \& Tombleson (2015) and Smith (2016, 2019)) 
concluded that both LBVs and WRs could be the products of binary evolution, LBVs being mass-gainers, and WRs -- mass-donors,  or stripped components. This idea was critically discussed by  \citet{Humphreys2016} and  \citet{Davidson2016}. Now we may confidently conclude only that we do not have enough data for statistical analysis. Modern evolutionary codes allow detailed calculations of binary system evolution (see for example \citet{Gotberg2018, Laplace2021}), thus making it possible to compare theoretical predictions for stripped stars with properties of real WR stars \citep{Gotberg2018}. Our current understanding is that binary interaction scenario cannot properly describe all observed WR stars, but most probably takes place in a part of them (see \citet{Shenar2024} and references therein).

Improving the statistics of binarity among WR stars is a crucial task. Among the first researchers who started to work on it was Virpi~S. Niemel{\"a}, who determined the stellar masses in binary systems and discovered and analyzed the spectroscopic orbits of many binary systems with WR components (see for example her works \citet{Niemela1995, Niemela2001}). Nowadays this work is being continued, both through spectroscopy \citep{Chene2022} and using high-resolution imaging \citep{Deshmukh2024}.

%%%%%%%%%%%%%%%%%%%%%%%%%%%%%%%%%%%%%%%%%%%%%%%
\section{Current problems}
\label{sec:problems}

\subsection{Search for new WR stars}

In our Galaxy there are 679 WR stars currently discovered\footnote{Galactic Wolf-Rayet Catalogue accessible at \url{https://pacrowther.staff.shef.ac.uk/WRcat/l}} 
\citep{Rosslowe2015}, while the theory based on current Milky Way star formation rate and duration of WR phase predicts the numbers of WR stars $\sim1200$ \citep{Rosslowe2015}. It means that significant part of WR stars are still not discovered. 
WR stars in the Galaxy align with spiral arms and the regions of star formation. Thus, the discovery of new WR stars through optical observations has been significantly limited due to the presence of dust extinction. Although the probability of finding a WR star during optical spectroscopic surveys remains \citep{MaizApellaniz2016,WeiZhang}, the majority of discoveries of new WR stars now happens through the observations in IR range. For example, \citet{2011Redeyes} identified 60 Galactic WR stars: candidates were selected using the photometry from {\it Spitzer} and Two Micron All Sky Survey (2MASS) surveys and confirmed by near-IR spectroscopy.   \citet{2012Shara} expanded the list by adding 71 more stars, also preliminary selected from  $J$ and $K$ band 2MASS photometry.

Moreover, the search for new WR stars is complicated by the effect of spectral mimicry -- low mass [WR] objects masquerading  as classical WR stars. Thanks to the results of {\it Gaia} mission \citep{Gaia2016} we now have reliable estimations of distances, and it helps to quickly understand the nature of individual objects and separate low-mass evolved stars from massive stars, and, moreover, to find the objects of unusual origin among low mass stars. For example, based on {\it Gaia} distance estimation  \citet{Gvaramadze2019} demonstrated that IRAS\,00500+6713 is a product of merging of two white dwarfs, while the star has WO-type spectrum. Counterexample is PMR\,5 star classified as [WN6] by  \citet{Morgan2003} that was recently reclassified as a classical WN. 

%Twenty-five objects in M33 have been identified as Woff--Rayet stars on the basis of narrow-band interference-filter photography. 

Systematic search for WR stars in nearby galaxies is ongoing since 1970-s. 
In 1972  first catalog of WR candidate stars located in M33 galaxy was published \citep{WrayCorso1972}. Twenty-five objects were selected based on photometry in three narrow-band filters (one filter is centered on He\,{\footnotesize II} 4686 -- the strongest emission line in WN-type WRs, one is centered on C\,{\footnotesize III}-{\footnotesize IV}~$\approx4650$ -- the strongest line in WC-type WRs, and one is centered on continuum).  This technique of selection of WR candidates for following spectroscopy has proven itself successfully and is still widely used \citep{MasseyConti1983,MasseyJohnson1998, NeugentMassey2011, Neugent2012}. As of today,  211 WR are known in M33 galaxy \citep{Massey2016} and 173 are in M31  \citep{Neugent2023}. Search for new WR and their following investigations are important  for understanding the impact of metallicity on mass-loss rate. 

%\subsection{Binarity of WR stars}
%%%%%%%%%%%%%%%%%%%%%%%%%%%%
\begin{figure}
	%\centering
	\centerline{
    \resizebox*{0.5\linewidth}{!}{\includegraphics%[angle=0,viewport=15 25 750 750,clip]
    {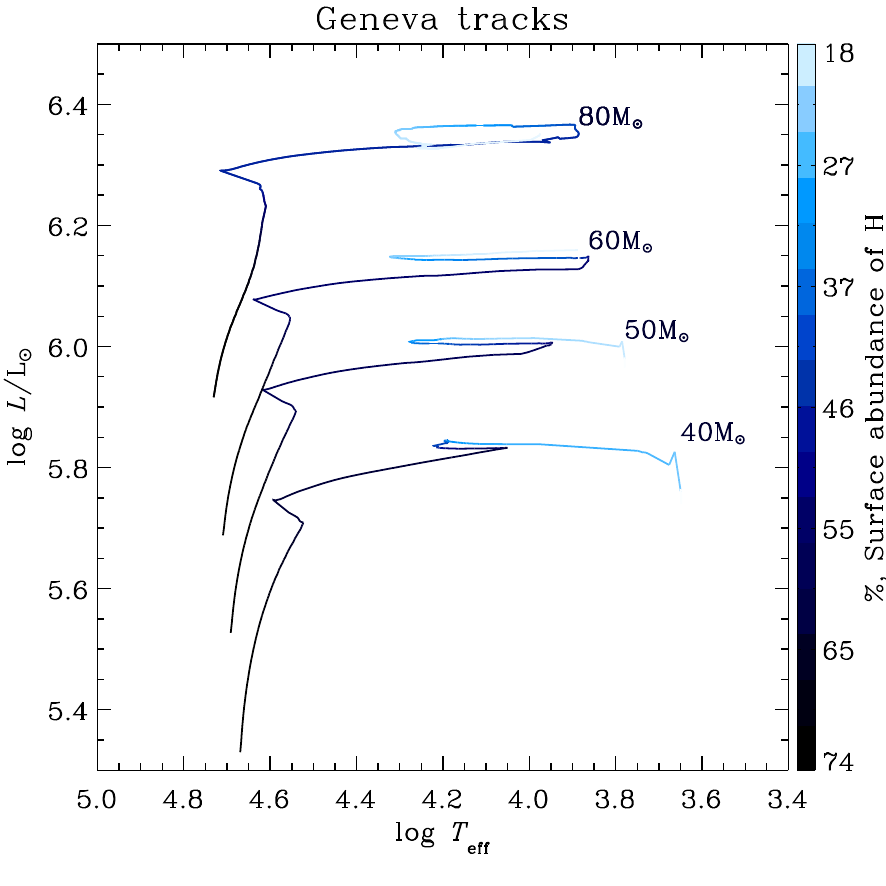}}
    \resizebox*{0.5\linewidth}{!}{\includegraphics%[angle=0,viewport=15 30 750 750,clip]
    {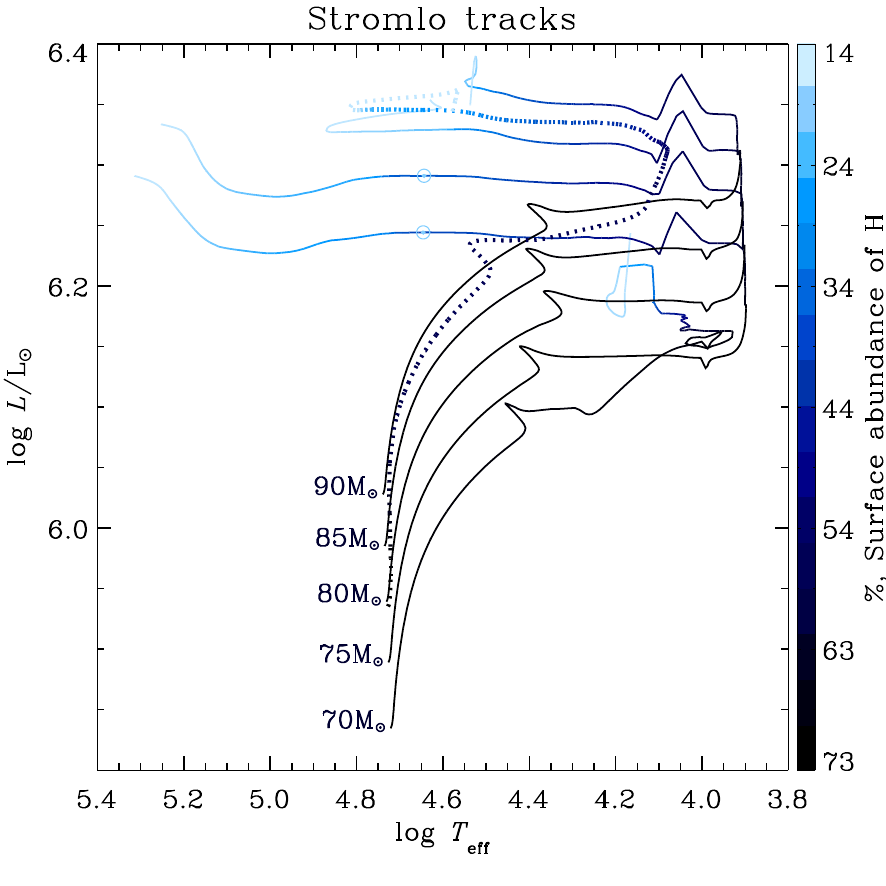}}}
\caption{HR diagram and evolutionary tracks for the low metallicity massive stars ($Z=0.002$) from  \citet{Georgy2013} (left panel) and \citet{Grasha2021} (right panel) evolutionary models. Geneva tracks do not produce massive WR stars at such low  metallicities, while Stromlo tracks show rather different evolutionary path and produce WR stars with $M_*>70\Msun$.}
\label{track}
\end{figure}
%%%%%%%%%%%%%%%%%%%%%%%%%%%%%%%%%%
%%%%%%%%%%%%%%%%%%%%%%%%%%%%%%
\subsection{Wolf-Rayet stars in low metallicity environment}

As the opacity of stellar wind depends on the metallicity, mass-loss rates of massive stars decrease with $Z$. It means that in low metallicity environment single massive stars are unable to lose enough material through the stellar wind to become WR star. Thus, the mass loss on pre-WR stages is of fundamental importance for the final fate of massive stars,
% and their chemical yields, 
and may influence e.g. the formation of long-duration Gamma Ray Bursts (GRBs) or the yields from early stellar generations.

% affects not only the microphysics of the stellar wind, but also 
Metallicity  impacts overall evolutionary behaviour of massive stars in a complicated way.  \citet{Georgy2013} calculated evolutionary tracks for stars in a wide range of initial masses at low metallicity. According to their calculations, metal-poor stars become colder after the end of hydrogen burning in the core and move along the track to the right side of the diagram. 
Unlike solar metallicity stars, metal-poor ones no longer return to the left side of the diagram (see Figure~\ref{track}), even when the effects of stellar rotation are taken into account in the models.
On the other hand, evolutionary tracks behave differently if the models implement Galactic Concordance abundances instead of solar, scaled-solar, or alpha-element enhanced abundances \citep{Grasha2021} -- the stars with the initial mass of $M_*>70\Msun$ with rotation $V/V_{\rm crit} = 0.2$ after the main sequence move to the right and then return to the left to the WR stars region, in contrast to predictions of  \citet{Georgy2013}.

 \citet{Yarovova2023} used evolutionary tracks from  \citet{Georgy2013} and   \citet{Grasha2021} for the study of emission-line stellar-like object in the nearby low-metallicity ($Z\sim0.1 Z_\odot$) dwarf galaxy NGC\,4068. They found the best agreement between the modelled and observed spectra for the model assuming ionization by low-metallicity WR star of $M_*\approx80\Msun$ mass, ionizing the nebula through the strong wind and enriching the interstellar medium with nitrogen. Thus, parameters of the object favor   \citet{Grasha2021} predictions. 
Therefore, understanding of mass loss rate requires complex approach that includes both studies of massive stars in different metallicites, and studies of present-day chemical abundances.

%%%%%%%%%%%%%%%%%%%%%%%%%%%%
\begin{figure}
	\centering
	\includegraphics[width=0.9\linewidth]{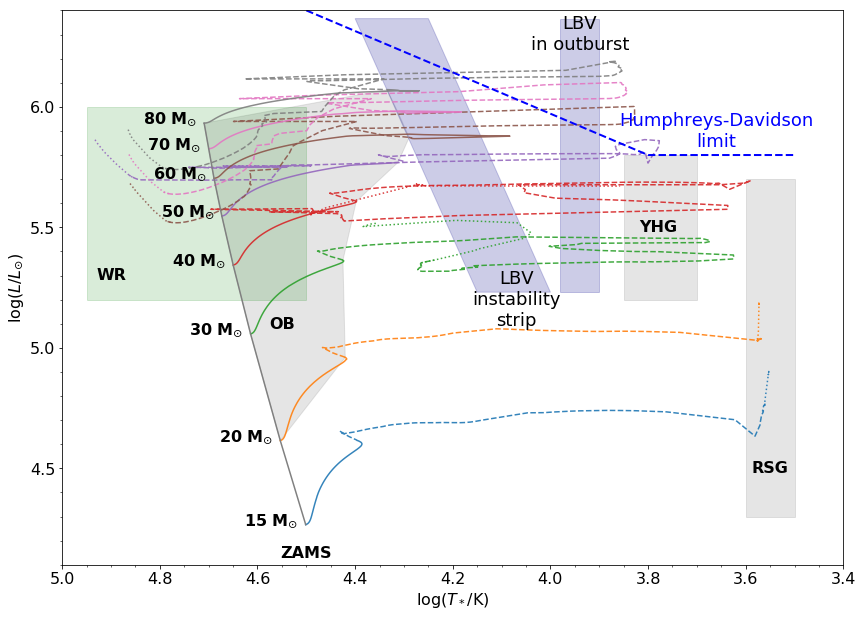}
	\caption{Hertzsprung--Russell (HR) diagram (luminosities versus effective temperatures). Color horizontal lines show evolutionary tracks, with solid parts showing the phase when hydrogen burns in the stellar core. The tracks are taken from Ekstr{\"o}m et al. (2012) \citep{Ekstrom2012}. ZAMS is zero age Main Sequence.}
	\label{HR}
\end{figure}
%%%%%%%%%%%%%%%%%%%%%%%%%%%%%%%%%%
\subsection{Wolf--Rayet stars  and red supergiants}

Isolated massive stars with initial masses in $8~\Msun\lesssim M_*\lesssim 20~\Msun$ interval undergo the transition to Red Supergiants (RSG) after hydrogen in the core is exhausted. For them RSG is the final stage before Type-II supernova (SNII) explosion. More massive stars ($40~\Msun \lesssim M_*< 90~\Msun$) evolve through LBV phase before reaching WR phase, and finally explode as SN Type~Ib/c.
The evolution of stars with $25~\Msun \lesssim M_*\lesssim 35~\Msun$ initial masses is more complicated. Observed population of supernovae Type~II progentiors in the local Universe shows that they are produced by the stars with initial masses $M_*\lesssim18\Msun$ \citep{Smartt2015}, and it suggests that more massive stars should either directly collapse to black holes, or evolve leftwards in HR diagram, thus becoming Wolf-Rayet stars. It is supported by the  luminosity distributions of cool supergiants in  Large and Small Magellanic Clouds \citep{Davies2018} that overlaps with those of apparently-single WR stars in both galaxies, thus suggesting a changing evolutionary sequence of massive stars with increasing initial mass.   %(Davies et al. 2018). 
At the same time, between the RSG and WR regions in HR diagram lies the region of {\it low luminosity LBVs} and yellow hypergiants (YHGs) (Figure~\ref{HR}). Therefore, the evolution of $25~\Msun \lesssim M_*\lesssim 35~\Msun$ initial mass may look like that:
 $$O \rightarrow RSG  \rightarrow YHG(?) \rightarrow low~LBV \rightarrow WR~~~(for~M_*\approx25-35~\Msun) $$

It means that the star with initial mass about $30~\Msun$ should pass through all four stages -- do such objects really exist?

\citet{Groh2013} demonstrated that the stars with  $20-25~\Msun$ initial masses are unstable on {\it low luminosity LBV} stage, and explode as core-collapse supernovae (SN Type~IIL/b). Prior to the explosion they show the spectrum similar to S\,Doradus during its minimum of brightness  (WN11h spectral type). 
\citet{Maryeva2020} have recently found a Galactic object, Wray\,15-906, with characteristics typical of a low luminosity LBV  and the spectrum looking like WN11h. 
It was identified via the detection of its infrared circumstellar shell (of $\approx2$~pc in diameter) with the Wide-field Infrared Survey Explorer (WISE) and the Herschel Space Observatory.  \citet{Maryeva2020} found that Wray\,15-906 is a relatively low-luminosity, $\log(L*/\Lsun)\approx5.4$, star with a temperature of $25\pm2$~kK, with position in HR diagram  corresponding to post-red supergiant with initial mass of $\approx25\Msun$.
Its spectrum and properties are consistent with theoretical predictions of  \citet{Groh2013}. Thus, the monitoring of this object, and the search for similar ones, may improve our understanding of the link between WRs and RSGs.

\section{Conclusion}

%Nowadays we understand quite well what Wolf-Rayet stars are, and they are no longer a hot topic in astrophysics. However they still are keystones for many directions of research, from theory of stellar atmospheres up to stellar populations in other galaxies and gravitational waves. Through the existence of  the Wolf-Rayet phenomenon WR stars are linked, besides massive stars, with low mass evolved stars. 

% В представленном обзоре собраны основные этапы исследования WR звёзд, показано что  понимания природы WR звёзд развивалось вместе с атомной физикой, численными методами, в сильной зависимости от инструментальных возможностей. 

Wolf-Rayet stars were discovered more than 150 years ago, and since then they were targets of numerous studies that combined astrophysical insights with advancements in atomic physics, development of numerical methods, and instrumental progress.
Nowadays we understand quite well what Wolf-Rayet stars are, and they are no longer a hot topic in astrophysics. However they still are keystones for many directions of research, from theory of stellar atmospheres up to stellar populations in other galaxies and gravitational waves. Through the existence of  the Wolf-Rayet phenomenon WR stars are linked, besides massive stars, with low mass evolved stars. 

We expect new  progress in understanding the Wolf-Rayet stars and mass-loss of massive stars in general from the upcoming advances in the instrumentation. For example, SITELLE imaging Fourier transform spectrometer  at Canada-France-Hawaii telescope \citep{Drissen2019} %(CFHT)
may provide spatially resolved line ratios and kinematics of WR nebulae 
thus hinting us on ionizing flux from the stars, and their mass loss history. James Webb Space Telescope will allow studying dust shells around WR stars \citep{Lau2022}, and to assess their role in enrichment of interstellar medium with organic compounds and carbonaceous dust. 
Spectropolarimetric surveys of WR and monitoring of variability of  linear polarization are also still topical. 

%The observations indicate that dust-forming carbon-rich Wolf–Rayet binaries can enrich the interstellar medium with organic compounds and carbonaceous dust.
%for spectroscopy and especially polarimetry that should improve our 

%SITELLE – imaging Fourier transform spectrometer –Canada-France-Hawaii telescope (CFHT).

%Spatially resolved line ratios and kinematics of WR nebulae provide information on their ionizing stars and mass loss history.

%Spectropolarimetric surveys of WR and monitoring of variability of  linear polarization are still actual

%Thus the understanding of WR stars is crucial for, and was historically developing along with, the atomic physics, numerical methods, and instrumental advancements.

%Обзор представляет историю исследования WR звёзд, показывает как 

\bigskip

{\bf Acknowledgements ~} I would like to thank the organizers of the conference for their kind invitation to present this overview.  This research received funding from the European Union's Framework Programme for Research and Innovation Horizon 2020 (2014--2020) under the Marie Sk\l{}odowska-Curie Grant Agreement No. 823734  (POEMS project). The~Astronomical Institute in Ond\v{r}ejov is supported by the project RVO:67985815.

%\clearpage 
%===========================================================================

\bibliography{WRbib}

}

\end{document}